\documentclass[sigplan]{acmart}
\renewcommand\footnotetextcopyrightpermission[1]{}
\usepackage[utf8]{inputenc}

\setcopyright{none}
\usepackage[normalem]{ulem}
\usepackage{epsfig, endnotes}
\usepackage{xcolor}
\usepackage{amsthm}
\usepackage{subfig}
\usepackage{xspace}
\usepackage{relsize}
\usepackage[htt]{hyphenat}
\usepackage{enumitem}
\setlist{nolistsep}
\usepackage{tablefootnote}
\usepackage{footnote}
\makesavenoteenv{tabular}
\usepackage{threeparttable}
\definecolor{ForestGreen}{RGB}{34,139,34}

\usepackage{pifont} % for the checkmark

\newcommand{\cmark}{\color{ForestGreen}{\small\ding{51}}}%
\newcommand{\xmark}{\color{red}{\small\ding{55}}}%

\def\ifmonospace{\ifdim\fontdimen3\font=0pt }
\def\C++{%
\ifmonospace%
    \C++%
\else%
    C\kern-.1167em\raise.30ex\hbox{\smaller{++}}%
\fi%
\spacefactor1000 }

\newcommand{\coolname}{\mbox{\textsc{Garmr}}} %\mbox prevents line breaks in the name

\usepackage{makecell}
\usepackage{multirow}
\usepackage{tikz}
\usepackage{pgfplots}
\usepackage{pgfplotstable}
\usepackage[nameinlink]{cleveref}

\newcounter{avcnt}

\usepackage{amsthm}

\PassOptionsToPackage{hyphens}{url}\usepackage{hyperref}

\begin{document}
\sloppy

%don't want date printed
\date{}

\title{\coolname{}: Defending the gates of PKU-based sandboxing}

%for single author (just remove % characters)
%\date{}
\settopmatter{printfolios=true}
%\maketitle

%\thispagestyle{empty}

\author{Alexios Voulimeneas}
\affiliation{
	\institution{imec-DistriNet, KU Leuven}
	\country{Belgium}
}
\email{alex.voulimeneas@kuleuven.be}

\author{Jonas Vinck}
\affiliation{
	\institution{imec-DistriNet, KU Leuven}
	\country{Belgium}
}
\email{jonas.vinck@kuleuven.be}

\author{Ruben Mechelinck}
\affiliation{
	\institution{imec-DistriNet, KU Leuven}
	\country{Belgium}
}
\email{ruben.mechelinck@kuleuven.be}

\author{Stijn Volckaert}
\affiliation{
	\institution{imec-DistriNet, KU Leuven}
	\country{Belgium}
}
\email{stijn.volckaert@kuleuven.be}

%%%%%%%%%%%%%%%%%%%%%%%%%%%%%%%%%%%%%%%%%%%%%%%%%%%%%%%%%%%%%%%%%%%%%%%%%%%%%%%%
%\subsection*{Abstract}
\begin{abstract}

Memory Protection Keys for Userspace (PKU) is a recent hardware feature that
allows programs to assign virtual memory pages to protection domains, and to
change domain access permissions using inexpensive, unprivileged
instructions. Several in-process memory isolation approaches leverage this
feature to prevent untrusted code from accessing sensitive program state and data. Typically, PKU-based isolation schemes need to be used in conjunction with mitigations such as CFI because untrusted code, when compromised, can otherwise bypass the PKU access permissions using unprivileged instructions or operating system APIs.

Recently, researchers proposed fully self-contained PKU-based memory isolation
schemes that do not rely on other mitigations. These systems use exploit-proof
call gates to transfer control between trusted and untrusted code, as well as a
sandbox that prevents tampering with the PKU infrastructure from untrusted code.

In this paper, we show that these solutions are not complete. We first develop
two proof-of-concept attacks against a state-of-the-art PKU-based memory
isolation scheme. We then present \coolname{}, a PKU-based sandboxing framework that can
overcome limitations of existing sandboxes. We apply \coolname{} to several
memory isolation schemes and show that it is practical, efficient and secure.

\end{abstract}

\maketitle

%-------------------------------------------------------------------------------
\section{Introduction}
%-------------------------------------------------------------------------------

Many computer programs contain untrusted components that must be isolated from
trusted components to guarantee the confidentiality and integrity of sensitive
program state or data. Modern operating systems provide the necessary isolation
only at the process boundary. This forces software developers to run components
as separate processes (sometimes referred to as compartments) that each have
their own virtual address space. One of the drawbacks of process-level
compartmentalization is that synchronous interaction between isolated components
incurs high performance overhead due to the expensive context-switching required
for inter-process communication. The research community has, therefore, proposed
several alternative forms of compartmentalization that have better performance
characteristics and are often more practical to apply to existing programs. The
core idea behind many of these techniques is to isolate untrusted components
in-process, thereby allowing them to share typical per-process resources with
the rest of the program~\cite{chen2016shreds, litton2016lwc,
  10.5555/1387589.1387611, vahldiek2018erim, hedayati2019hodor, hsu2016smv,
  tarkhani2020utiles, 9152671, burow2019sok, DBLP:conf/ndss/ParkDGNVF20,
  xom-switch, 10.1145/3445814.3446728, 9152611, 10.1145/3064176.3064217,
  255298, gravani2021iskios, park2018libmpk}.

Memory Protection Keys for Userspace (PKU) is a hardware feature that is
available on recent Intel and AMD server and desktop CPUs~\cite{intel-mpk}. PKU
allows programs to be compartmentalized by assigning memory pages to memory
protection domains whose access permissions can be set individually by modifying
the content of the PKU control register (\texttt{PKRU}). PKU exposes a set of
unprivileged instructions to read and modify said register. This allows a
program to quickly disable access to all compartments that must be isolated from
the currently executing compartment, without having to pay the high cost of the
system calls and TLB invalidations required to change page permissions through
conventional means. However, an attacker can exploit vulnerabilities, hijack the
control-flow of a program and abuse \texttt{PKRU}-updating instructions to
modify \texttt{PKRU}, enabling access to any compartment's memory.

Recently, researchers proposed ERIM~\cite{vahldiek2018erim} and
Hodor~\cite{hedayati2019hodor}, two efficient PKU-based memory isolation schemes
that isolate an application's trusted from its untrusted components by placing
them in different memory protection domains. Both systems have built-in
sandboxes that prevent adversaries from bypassing the isolation scheme by
executing unsafe instructions that modify the \texttt{PKRU} register. Similar to
previous work, we refer to these sandboxes as \emph{PKU-based
  sandboxes}~\cite{connor2020pkupitfalls}. Unfortunately, these sandbox
implementations have known security and usability problems. ERIM, for example,
relies on static binary instrumentation (SBI) to neutralize any unsafe
instructions in the protected program. However, as SBI cannot reliably
distinguish code from data, ERIM could leave some unsafe instructions
untouched~\cite{10.1007/978-3-642-23808-6_34}. Currently, ERIM's sandbox also
marks pages that contain unsafe instructions as non-executable, which could lead
to usability issues~\cite{vahldiek2018erim}.  Hodor's sandbox uses hardware
breakpoints to ensure the program cannot execute unsafe instructions. This
approach does not rely on SBI like ERIM's, but both systems can still be
bypassed using the kernel as a confused deputy~\cite{connor2020pkupitfalls}.

In addition to the aforementioned problems, we found two more security flaws in
the design of Hodor's sandbox and managed to exploit these flaws in new
proof-of-concept attacks that we present in this paper. We then designed and
implemented \coolname{}, a new PKU-based sandboxing framework that can protect
PKU-based memory isolation schemes against all known \texttt{PKRU}-tampering
attacks (except for the signal context attacks described
in~\Cref{sec:signals}). We used \coolname{} to develop sandboxes for two
existing isolation schemes and evaluated the performance and security of the
resulting systems. We conclude that \coolname{} enables practical, efficient,
and secure PKU-based sandboxing.

In summary, our paper contributes the following:

\begin{itemize}
	\item We identified new design flaws in Hodor's PKU-based sandbox and
      developed two new proof-of-concept attacks that exploit these
      flaws~\cite{hedayati2019hodor}.

	\item We present \coolname{}, a new PKU-based sandboxing framework, and
      apply our framework to develop sandboxes for two state-of the art
      PKU-based memory isolation schemes: ERIM~\cite{vahldiek2018erim} and
      XOM-Switch~\cite{xom-switch}. The resulting sandboxes stop all known
      attacks (except for the signal context attacks described
      in~\Cref{sec:signals}), including the new attacks we present in this
      paper~\cite{vahldiek2018erim, hedayati2019hodor, connor2020pkupitfalls}.
	
	\item We perform an extensive evaluation of the constructed sandboxes on
      real-world server applications and show that \coolname{} enables
      practical, efficient, and secure PKU-based sandboxing.
\end{itemize}

%-------------------------------------------------------------------------------
\section{Background}
%-------------------------------------------------------------------------------
\label{sec:background}

PKU utilizes a new user-mode register (\texttt{PKRU}) to control access
rights\footnote{The \texttt{PKRU} register only controls data accesses,
  instruction fetches are not similarly restricted.}  to memory pages that are
tagged with one of 16 available protection keys. The \texttt{PKRU} register is
32 bits wide and has two bits (access disable and write disable) for each
key. These bits are checked during memory accesses for all the pages that are
associated with a key. The OS provides new system calls, \texttt{pkey\_alloc}
and \texttt{pkey\_free}, to allocate and free protection keys respectively. A
process can tag a page with a key by using the new \texttt{pkey\_mprotect}
system call and access the \texttt{PKRU} register with unprivileged
instructions; \texttt{rdpkru} for read and \texttt{wrpkru} for write
accesses. The \texttt{xrstor} instruction can also update the \texttt{PKRU}
register if bit 9 in the \texttt{eax} register is set prior to the instruction
execution.

\subsection{PKU-based Memory Isolation Schemes}

Some memory isolation schemes leverage PKU to isolate trusted from untrusted
components~\cite{hedayati2019hodor, vahldiek2018erim,
  10.1145/3445814.3446728}. These systems typically tag memory pages containing
trusted code and data with a different protection key than pages containing
untrusted code, thereby placing them in different memory protection
domains. Trusted code is then allowed to access both domains, whereas untrusted
code can only access the untrusted domain. Researchers have also used PKU to
harden JavaScript engines~\cite{DBLP:conf/ndss/ParkDGNVF20}, reinforce other
exploit mitigations~\cite{10.1145/3064176.3064217, burow2019sok,
  gravani2021iskios}, and provide software abstractions for isolation and
sandboxing~\cite{park2018libmpk}.

One major challenge for PKU-based systems is to prevent attackers from tampering
with the \texttt{PKRU} register by exploiting a memory vulnerability in the
untrusted code and by subsequently locating and executing a
\texttt{PKRU}-modifying instruction. One way to prevent such attacks is to apply
exploit mitigations to the untrusted code~\cite{larsen.etal+14,
  abadi2005control, burow2017control}. However, these mitigations can introduce
non-trivial run-time overheads and often cannot fully prevent \texttt{PKRU}
tampering.

\begin{figure}[t]
	\centering
	\includegraphics[scale=0.35]{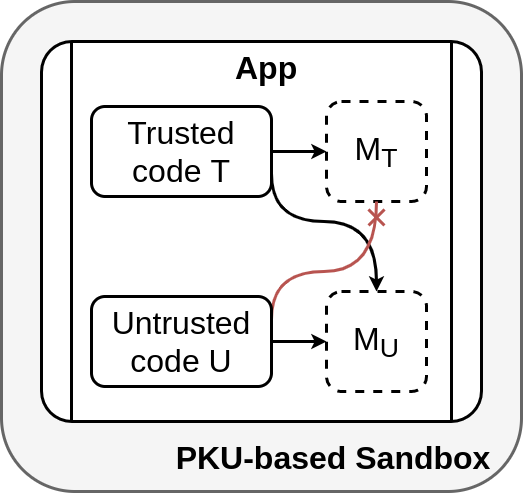}
	\caption{The goal of a PKU-based sandbox is to block an attacker that has
      seized control of U's execution from accessing M\textsubscript{T}, while
      permitting legitimate memory accesses from U to M\textsubscript{U} and
      from T to both M\textsubscript{T} and M\textsubscript{U}.}
	\label{fig:background}
\end{figure}

Hodor and ERIM are PKU-based isolation systems that do not rely on such
mitigations~\cite{hedayati2019hodor, vahldiek2018erim}. Both approaches use two
domains, M\textsubscript{T} and M\textsubscript{U}, that contain trusted and
untrusted components respectively. Transferring control from trusted components
to untrusted components or vice versa happens via well-defined, exploit-proof
instruction sequences also known as \emph{call gates}. Hodor and ERIM each use a
sandbox to prevent attackers that manage to compromise an untrusted component
from accessing M\textsubscript{T} by abusing \texttt{PKRU}-modifying
instructions (\texttt{wrpkru}, \texttt{xrstor}) or system calls like
\texttt{pkey\_mprotect} as depicted in \autoref{fig:background}.

\subsection{A Closer Look at ERIM}
\label{sec:erim}

ERIM compartmentalizes applications through binary rewriting. It assumes the
trusted components T are not exploitable, but does not make any assumptions
about the untrusted components U. ERIM implements call gates using so-called
\emph{safe instructions}. Safe \texttt{wrpkru} instructions are those that are
immediately followed by either instructions to validate \texttt{PKRU}'s state at
run time (ensuring that M\textsubscript{T} is locked by \texttt{PKRU}) or by a
jump to T. Safe \texttt{xrstor} instructions, on the other hand, are immediately
followed by instructions that check if bit 9 of eax is set. If one of these
run-time validations fails, the control flow jumps to an instruction sequence
that terminates the application. Otherwise, the program execution continues. We
refer to any other \texttt{wrpkru} and \texttt{xrstor} instructions as
\emph{unsafe instructions} similar to previous
work~\cite{connor2020pkupitfalls}. ERIM's call gates are not exploitable, since
they do not contain unsafe instructions. However, an attacker that controls U
could abuse unsafe instructions found outside call gates to change
\texttt{PKRU}, thereby allowing U to access M\textsubscript{T}. Attackers can
easily find unsafe instructions because (a) they could appear as operands of
other instructions, and (b) x86 instructions do not have a fixed size and the
CPU, therefore, allows programs to execute instruction operands as if they were
regular instructions themselves.

At startup time, ERIM’s PKU-based sandbox scans all executable pages of the
protected application using the \texttt{/proc/self/mem} interface to verify the absence of
exploitable unsafe instructions in M\textsubscript{U} pages. Any executable page
that contains unsafe instructions is marked as non-executable. Consequently, any
attempt to execute code from such a page will trigger a fault that is handled by
ERIM’s sandbox, and the sandbox terminates the program prematurely. To prevent
this, ERIM first uses a static binary rewriter to replace instruction sequences
that contain unsafe \texttt{wrpkrus} and \texttt{xrstors} by functionally
equivalent sequences that do not contain such unsafe occurrences.  At run time,
the sandbox intercepts and monitors \texttt{mmap}, \texttt{mprotect} and
\texttt{pkey\_mprotect} system calls from U that can introduce unsafe
instructions or allow access to M\textsubscript{T}. ERIM provides two sandbox
implementations: one version that is based on
\texttt{ptrace}~\cite{intel-manuals}, and a more efficient one that requires
minor kernel modifications.

\subsection{A Closer Look at Hodor}
\label{sec:hodor}

Hodor partitions applications into trusted and untrusted libraries.
%% \footnote{Hodor refers to its own call gates
%%   as \emph{trampolines}. As described in the original
%%   paper~\cite{hedayati2019hodor}, Hodor employs the standard function
%%   call/return interface but interposes a trampoline on each call to change the
%%   view of the address space to that of the library being entered.}.
Hodor's trusted application loader ensures that untrusted libraries U can only
interact with trusted libraries T through call gates similar to ERIM's. However,
unlike ERIM, Hodor does not rely on code rewriting.

Hodor's PKU-based sandbox monitors the application at run time to stop an
attacker from abusing unsafe \texttt{wrpkru} instructions that change
\texttt{PKRU}. When an application attempts to mark a page as executable, the
trusted loader first scans the page for unsafe \texttt{wrpkru} instructions and
marks the page as non-executable if it contains unsafe instructions. Attempts to
execute code from such a page trigger a fault. Upon receiving that fault,
Hodor's modified OS kernel attempts to put hardware breakpoints on all unsafe
instructions on the same page, and it marks the page as executable. If the page
contains more unsafe \texttt{wrpkru} instructions than the maximum number of
hardware breakpoints the CPU supports, Hodor will single-step through the page
instead~\cite{intel-manuals}. This mechanism ensures that all unsafe
\texttt{wrpkru} instructions will be vetted by Hodor's kernel. When a hardware
breakpoint is triggered, Hodor's modified kernel terminates the program
execution. Hodor can reclaim hardware breakpoint slots in the debug registers if
and when necessary. However, when doing so, it will always mark the pages these
breakpoints point to as non-executable.

%% To reach its goals, the trusted loader intercepts
%% and monitors \texttt{mmap}, \texttt{mprotect} and \texttt{pkey\_mprotect} system
%% calls.

\subsection{XOM-Switch}
\label{sec:xom}

PKU can also be used for eXecute-Only Memory (XOM). Recent kernels (Linux 4.9+)
support XOM through an enhanced version of the \texttt{mprotect} system
call\footnote{The kernel assigns keys to memory pages and updates the
  \texttt{PKRU} register accordingly. No user space code is involved.}. Support
for XOM is missing in \texttt{libc} and compilers,
though. XOM-Switch~\cite{xom-switch} patches the dynamic loader/linker and
\texttt{libc} to apply XOM memory to ELF binaries. However, the defense is
limited in power unless it is combined with an additional mitigation such as
CFI, since an attacker can abuse PKU to disable XOM. The authors of XOM-Switch
acknowledged its limitations and proposed the use of Intel’s Control-flow
Enforcement Technology (CET)~\cite{intel-manuals}, an upcoming feature of Intel
CPUs, as the ultimate solution against such attackers.

%-------------------------------------------------------------------------------
\section{Threat Model}
%-------------------------------------------------------------------------------

For this paper, we make the following assumptions about the host system, the targeted application and the attacker. Our assumptions are in line with work in the area~\cite{connor2020pkupitfalls, vahldiek2018erim, hedayati2019hodor}:

\begin{itemize}
	
\item \textbf{Host System.} We assume Protection Keys for Userspace (PKU)~\cite{intel-mpk} to be available on the target platform and we trust its implementation. The kernel is also considered part of the Trusted Computing Base (TCB).

\item \textbf{Targeted Application.} We do not make any assumptions about the untrusted code U, but similar to previous work~\cite{connor2020pkupitfalls, vahldiek2018erim, hedayati2019hodor}, we assume that the initial state of the targeted application is not compromised and that the PKU-based sandbox is initialized correctly. The trusted code T of the application, however, is considered free of exploitable bugs. We also assume for simplicity that there are only two levels of trust (T and U), and that the application is not using PKU for any purposes, except for memory isolation.

\item \textbf{Attacker.} We consider an attacker that controls U of the targeted application with the goal to access M\textsubscript{T}. For example, an adversary can use code-reuse and control-flow hijacking attacks to exploit unsafe instructions in executable pages of M\textsubscript{U} to tamper with \texttt{PKRU}'s state, enabling access to M\textsubscript{T}. Mitigations like software diversity~\cite{larsen.etal+14} and CFI~\cite{abadi2005control, burow2017control} raise the bar for such attacks, but we do not rely on such defenses. Attacks that target the underlying hardware such as transient execution~\cite{gens2017lazarus, lipp2018meltdown, kocher2019spectre} and remote-fault injection~\cite{seaborn2015exploiting, van2016drammer, brasser2017can, kenjar2020v0ltpwn} are considered out of scope for this paper. 

\end{itemize}
%-------------------------------------------------------------------------------
\section{Challenges}
%-------------------------------------------------------------------------------

ERIM and Hodor have built-in sandboxes to prevent attackers that control U from
accessing M\textsubscript{T} by any means. However, there are several security
and usability challenges associated with PKU-based
sandboxing~\cite{vahldiek2018erim, hedayati2019hodor, connor2020pkupitfalls}. We
describe these issues below, introduce two new attacks against Hodor, and also
discuss potential solutions.

\subsection{Handling of unsafe instructions}
\label{sec:inshandling}

ERIM and Hodor eliminate or detect unsafe instructions that can tamper with
\texttt{PKRU}'s state. We show below that the proposed techniques to handle
unsafe instructions lead to security and usability issues.

First, Hodor's sandbox ensures that unsafe \texttt{wrpkru} instructions are
vetted by its modified kernel, leveraging hardware breakpoints and single-step
execution. We carefully examined the open source implementation of
Hodor\footnote{https://github.com/hedayati/hodor} and discovered that it does
not monitor unsafe \texttt{xrstor} instructions. Therefore, an attacker that
controls U can abuse these unsafe \texttt{xrstor} instructions to unlock
M\textsubscript{T}.

ERIM, on the other hand, relies on SBI to neutralize unsafe \texttt{wrpkru} and
\texttt{xrstor} instructions. In addition, ERIM's sandbox inspects the program
at run time to ensure no new unsafe instructions are introduced in the
executable pages of M\textsubscript{U}. If U tries to map a page that contains
unsafe instructions, it is marked as non-executable by ERIM's sandbox to stop
attackers from exploiting them. We used the open source implementation of
ERIM\footnote{https://github.com/vahldiek/erim} and repeated the experiments
described in the original paper~\cite{vahldiek2018erim}. We verified that ERIM's
approach works for the tested system (Debian 8, Linux 4.9.60) and applications.

However, we could not replicate the experiments on recent systems (Ubuntu 18.04,
Linux 5.3.18). The reason is that ERIM's tested system is fairly old, and uses
outdated versions of binaries such as the dynamic linker \texttt{ld.so},
\texttt{libc.so}, and \texttt{libm.so} (version 2.19). ERIM's rewriting strategy
cannot eliminate all unsafe instructions in recent versions of these binaries
(version 2.27), either due to SBI's inability to distinguish code from
data~\cite{10.1007/978-3-642-23808-6_34} or limitations of the current
prototype. Furthermore, ERIM's sandbox marks pages containing unsafe
instructions as non-executable, creating usability issues as described in
Section 1 of ERIM~\cite{vahldiek2018erim}. Even trivial compartmentalized
programs are terminated early by ERIM's sandbox in recent systems, since they
attempt to execute code in pages that contain unsafe instructions. We can modify
the sandbox to not mark these pages as non-executable, but this would lead to
security issues, since an attacker that controls U can exploit unsafe
instructions from these pages.

\subsection{PKU Pitfalls}
\label{sec:pitfalls}

Conor et al.~\cite{connor2020pkupitfalls} developed proof-of-concept exploits
against ERIM and Hodor that bypass their PKU-based sandboxes. These attacks use
the kernel as a confused deputy, taking advantage of OS abstractions that are
agnostic of PKU-based memory isolation schemes. We briefly describe them below.

\subsubsection{Inconsistencies of PT and PKU permissions.}\par\hfill
\label{sec:inconsistencies}

The OS exposes system calls that do not \emph{respect} the enforced page table
(PT) and PKU permissions. An attacker can use \texttt{process\_vm\_readv},
\texttt{process\_vm\_writev} and \texttt{ptrace} system calls to directly access
M\textsubscript{T} from U. PKU-based sandboxes should intercept and monitor
these system calls to prevent such accesses. This introduces negligible
performance overhead, since these calls are rarely used, as shown in previous
work~\cite{connor2020pkupitfalls}.

Another method to circumvent enforced page table and PKU permissions is to use
the \texttt{procfs} interface. A process can open its \texttt{/proc/self/mem}
file, in which positions correspond to addresses in the process' address space,
and perform I/O operations on it. By design these operations ignore page table
and PKU permissions, allowing an attacker to directly access M\textsubscript{T}
from U, or modify non-writable code to tamper with \texttt{PKRU}'s
state. PKU-based sandboxes should at least intercept and monitor
\texttt{open}-like system calls to prevent such attacks. However, it is
shown that this adds huge overhead, unless an efficient system call interception
and monitoring mechanism is used~\cite{connor2020pkupitfalls}.

\subsubsection{Mappings with mutable backings.}\par\hfill
\label{sec:mappings}

More problems arise when mapped memory is backed by a mutable file. An attacker
can directly perform I/O operations on this file and modify it, regardless of
the page table and PKU permissions of the mappings that are backed by it. These
modifications are reflected to the corresponding mappings. The OS also allows
multiple mappings of the same shared memory with different page permissions that
refer to the same physical memory. Therefore, an attacker can modify an
immutable and executable mapping through another writable mapping of the same
shared memory. In both mentioned cases, the attacker can add unsafe instructions
to executable pages without being detected by the PKU-based sandbox.

\subsubsection{Changing code by relocation.}\par\hfill
\label{sec:relocation}

Attackers can also introduce unsafe instructions, without modifying executable
pages of M\textsubscript{U}. First, they can create two \emph{non-neighboring}
mappings that each contains part of an unsafe instruction at the page
boundary. The PKU-based sandbox scans the mappings for unsafe instructions, and
both are considered safe, since they do not contain \emph{complete} unsafe
instructions. Then attackers can use the \texttt{mremap} system call to move the
mappings next to each other, and \emph{form} an unsafe instruction. To stop this
attack, the sandbox should re-scan the page boundaries for unsafe instructions
after every relocation~\cite{connor2020pkupitfalls}.

\subsubsection{Influencing intra-process behavior with seccomp.}\par\hfill
\label{sec:influencing}

ERIM and Hodor use the new \texttt{pkey\_mprotect} system call to isolate
M\textsubscript{T} from U. The trusted code T allocates a dedicated memory
region (M\textsubscript{T}) to store secrets, e.g., encryption keys, and uses
\texttt{pkey\_mprotect} to associate it with a different protection domain than
U. However, a malicious \texttt{seccomp} filter can deny these calls and return
a success value, tricking T to store sensitive data in memory that is not
properly isolated from U. A sandbox can prevent this attack by intercepting and
restricting \texttt{prctl} and \texttt{seccomp} system
calls~\cite{connor2020pkupitfalls}.

\subsubsection{Modifying trusted mappings.}\par\hfill
\label{sec:modtrustmapping}

Attackers can also change the virtual address space to access isolated memory or
modify T. For example, an adversary can invoke a \texttt{pkey\_mprotect} system
call to modify the protection key that a page of M\textsubscript{T} is tagged
with. Hodor prevents such attacks by passing the addresses of T and
M\textsubscript{T} to the kernel, and denying any attempt to change them from
U~\cite{hedayati2019hodor}.

\subsubsection{Race conditions.}\par\hfill
\label{sec:races}

PKU-based sandboxes scan executable pages for unsafe instructions. An attacker
that controls multiple threads can exploit race conditions in the memory
scanning process to bypass these sandboxes and add unsafe instructions to
executable memory. To protect against such attacks, the sandbox should initially
mark the page as non-writable and non-executable, scan it, and then mark it as
executable in case that it does not contain unsafe
instructions~\cite{connor2020pkupitfalls}. Furthermore, ERIM's sandbox cannot
reliably determine trusted mappings (requested by T), allowing an attacker to
trick the sandbox to accept untrusted mappings (originated from U) as trusted
ones\footnote{This attack targets the \texttt{ptrace}-based implementation of
  ERIM's sandbox.}.

\subsubsection{Signal context attacks.}\par\hfill
\label{sec:signals}

An adversary can abuse signals to modify \texttt{PKRU}'s state. The CPU state is
exposed both during signal delivery, and the return from a signal handler. An
attacker can use system calls such as \texttt{sigreturn} and
\texttt{sigaltstack} to tamper with \texttt{PKRU}'s state directly or indirectly
without using \texttt{wrpkru} or \texttt{xrstor} instructions. For example, an
attacker can craft a CPU state on the stack and use the \texttt{sigreturn}
system call to restore an arbitrary value to the PKRU
register~\cite{connor2020pkupitfalls}.

\subsection{New PKU Pitfalls}
\label{sec:newpitfalls}

We examined the open source implementation of Hodor, and we developed two new
proof-of-concept attacks against it that bypass the memory isolation scheme and
are not stopped by its sandbox. Our attacks, specifically target Hodor's
instruction vetting mechanism that uses debug registers to monitor unsafe
instructions.

\subsubsection{Vetted unsafe instruction relocation.}\par\hfill
\label{sec:relocation2}

Hodor's modified kernel loads the addresses of unsafe instructions of an
executable page in debug registers, and terminates execution if a hardware
breakpoint is triggered. An attacker can relocate this page using
\texttt{mremap} system call. Hodor's sandbox, however, does not intercept this
call. Consequently, debug registers are not updated, allowing exploitation of
unsafe instructions in that specific page. \textbf{\emph{This is a different
    attack than the one described in~\Cref{sec:relocation}, and requires
    additional measures}}. To deal with this attack, Hodor's sandbox should
intercept \texttt{mremap} system call, mark the page as non-writable and
non-executable, relocate the page, update the debug registers to store the new
addresses of the unsafe instructions, and finally mark the relocated page as
non-writable and executable.

\subsubsection{Incomplete debug register update.}\par\hfill
\label{sec:incompleteupdate}

We also discovered a fundamental flaw in Hodor's approach to monitor unsafe
instructions. The key problem is that threads share code and data, while debug
registers are unique to each thread.  Hodor's sandbox vets unsafe instructions
using debug registers as described in~\Cref{sec:hodor}. However, when one thread
maps memory containing unsafe instructions, it is not possible to update debug
registers for all threads synchronously, unless a process-wide \emph{stop the
  world} mechanism is implemented. As a result, an attacker can use one thread
to introduce unsafe instructions, and abuse them from another thread to tamper
with \texttt{PKRU}'s state and unlock M\textsubscript{T} to U.

\begin{figure*}[t]
	\centering
	\includegraphics[width=13cm]{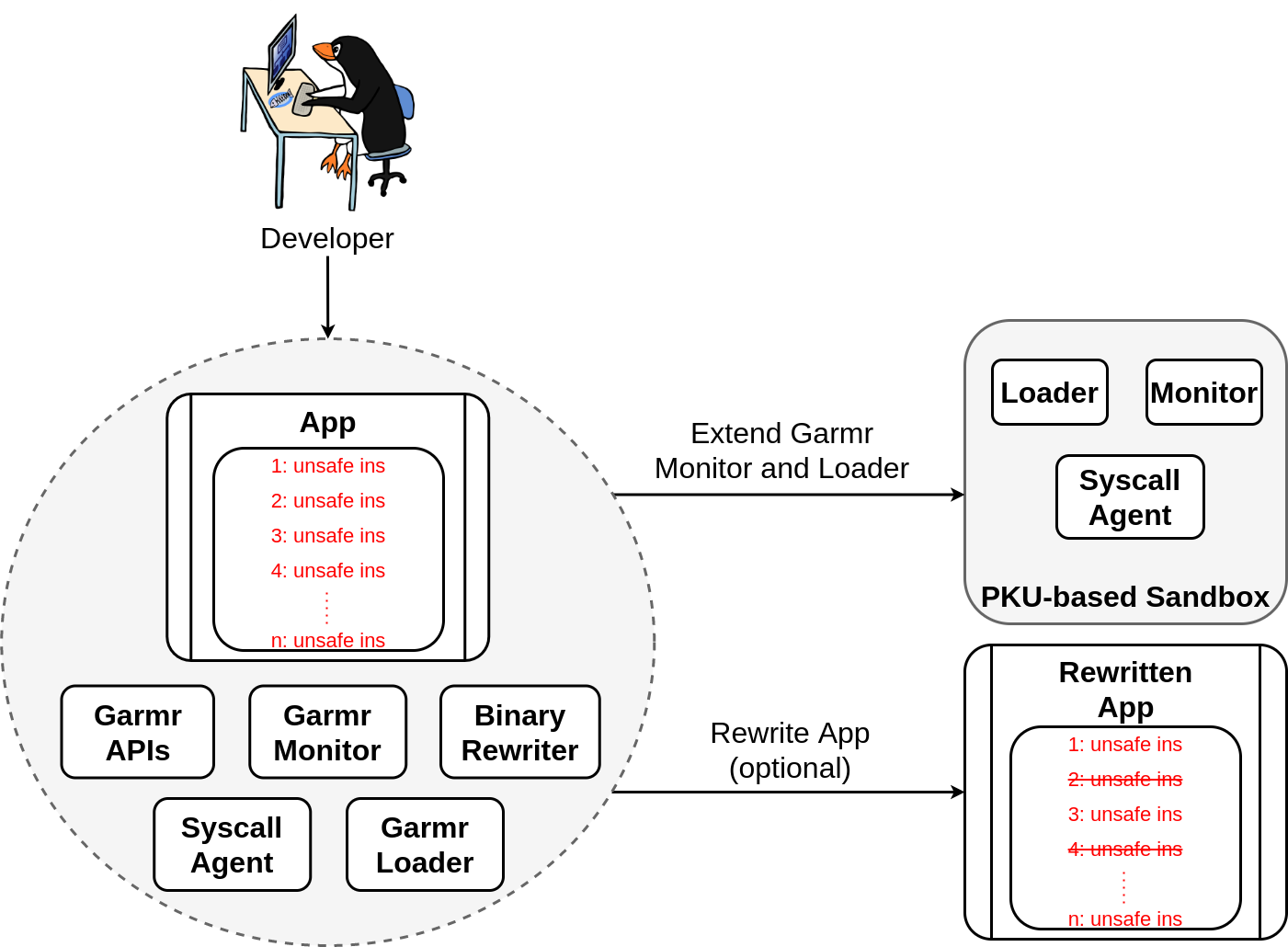}
	\caption{The developer (optionally) uses the SBI tool to eliminate as many
      unsafe instructions as possible from the applications. We represent the
      eliminated unsafe instructions with strike-through text in the rewritten
      application. Next, the developer uses \coolname{} APIs to extend the
      \coolname{} Loader and \coolname{} Monitor. The produced PKU-based sandbox
      consists of the modified loader, the modified monitor and the syscall
      agent.}
	\label{fig:workflow}
\end{figure*}

%-------------------------------------------------------------------------------
\section{Design, Workflow and Use Cases}
%-------------------------------------------------------------------------------
\subsection{Design}
\label{sec:design}

With the aforementioned challenges in mind, we designed and implemented \coolname{}, a general framework that developers can use to build PKU-based
sandboxes. \coolname{}'s design draws inspiration from the following three
observations:

\begin{enumerate}[label=\textbf{\arabic*.},ref={Observation~O\arabic*},noitemsep]

\item Instruction vetting techniques such as the ones used in Hodor do not need
  to be implemented in the kernel. Instead, we can implement them in a
  user-space tool using the \texttt{ptrace} API.

\item Static binary rewriting techniques such as those used in ERIM can
  eliminate many, but not necessarily all unsafe instructions in an application
  (see~\Cref{sec:inshandling}). Eliminating unsafe instructions is useful even
  for sandboxes that use hardware breakpoints for instruction vetting, because
  doing so reduces pressure on the processor's debug registers.

\item PKU-based sandboxes should intercept and monitor only a few system calls,
  and the vast majority of these calls are not frequently invoked by
  applications. System calls used to open file descriptors (e.g.,
  \texttt{sys\_open}) are an exception.
  
\end{enumerate}

Our framework consists of an SBI tool (Binary Rewriter), a \texttt{ptrace}-based
monitor (\coolname{} Monitor), a custom loader (\coolname{} Loader), a set of
APIs to extend the loader and the monitor (\coolname{} APIs), and a minimal
in-kernel syscall agent (Syscall Agent). We describe these components below.

\textbf{Binary Rewriter.} SBI cannot neutralize all unsafe instructions
(see~\Cref{sec:inshandling}). However, an SBI tool can still eliminate a portion of unsafe instructions from the binaries. Consequently, the PKU-based
sandbox has to monitor fewer unsafe instructions. Currently, we use ERIM's SBI
tool to rewrite binaries~\cite{vahldiek2018erim}.

\textbf{\coolname{} Loader.} We developed a custom application loader that, upon
startup, maps a \emph{special} executable and non-writable page containing an
\texttt{rdpkru} instruction into the application's address space. We use this
instruction to implement an interface for a \texttt{ptrace}-based monitor to
read \texttt{PKRU} register (see \textbf{\coolname{} APIs} below). Then, the
loader initializes the syscall agent through a \texttt{prctl} option that we
added to the kernel, and completes the program loading.

\textbf{\coolname{} Monitor.} We provide a \texttt{ptrace}-based monitor that
intercepts and monitors system calls. The monitor also injects the custom loader
into the application before launching it, and ensures that the access
permissions of the aforementioned special page are not changed and that it is
not unmapped. The monitor can (optionally) log the invoked system calls and
their results. \coolname{} spawns a new monitor for every newly created process
or thread.

\textbf{Syscall Agent.} \texttt{ptrace}-based monitors introduce high overhead
due to expensive context-switching and TLB invalidations. To avoid this cost, we
implemented a minimal in-kernel syscall agent that can let certain system calls
bypass the ptracer. Each thread has its own agent. Our custom loader initializes
the agent, providing a list of system calls (\texttt{SList}) and a list of
\texttt{inodes} (\texttt{IList}). The agent forwards to \coolname{} monitor
\emph{only} the system calls of \texttt{SList}, while the rest are executed
natively, bypassing \texttt{ptrace}. It also denies opening of \emph{sensitive}
files included in \texttt{IList}. This functionality can also be implemented
in a \texttt{ptrace}-based monitor like the \coolname{} monitor, but it would
add substantial run-time overhead, as shown in previous
work~\cite{connor2020pkupitfalls}.

\textbf{\coolname{} APIs.} \coolname{} implements generic macros and functions
that developers can use to extend the \coolname{} monitor and the \coolname{}
loader to implement PKU-based sandboxes, tailored to the needs of the memory
isolation scheme. We briefly describe the most important APIs. First, developers
can use simple macros to add/remove items to/from \texttt{SList} and
\texttt{IList}. In addition, \coolname{} provides macros and functions to write
system call handlers, and to access registers and memory. Developers can also
use APIs to scan the application's memory for unsafe instructions on
executable pages, similar to ERIM and Hodor.

Moreover, our framework provides a \texttt{ptrace}-based interface to read
\texttt{PKRU}, reliably identifying the current domain
permissions. Specifically, we use \texttt{ptrace} to change control flow and
execute the \texttt{rdpkru} instruction that the loader mapped into the address
space at startup. This functionality is important, since the current version of
the \texttt{ptrace} API cannot access the \texttt{PKRU}\footnote{We expect that
  to change in the near future.}.  Finally, \coolname{} exposes APIs to access
debug registers and to set/unset hardware breakpoints, an emulation engine for
x86 instructions and a rich logging infrastructure that can help sandbox
developers identify and fix bugs. We describe use cases for these APIs
in~\Cref{sec:usecases}.

\subsection{Development Workflow}

The first (optional) step of the development workflow is to eliminate a portion
of the unsafe instructions that are present in the to-be-sandboxed application
with an SBI tool. We rewrite the application binary, system libraries and other
dependencies. Then, the developer uses the provided APIs to extend the loader
and the monitor to build PKU-based sandboxes for different memory isolation
schemes. For example, the developer has the freedom to implement different
system call policies, define which instructions are safe and which are not, and
use debug registers to vet unsafe instructions.

Each produced PKU-based sandbox consists of the following components: a modified
loader, a modified monitor and the syscall agent. For simplicity, we refer to
the first two components as the loader and the monitor respectively in the rest
of the paper. \emph{\textbf{Note that \coolname{} does not modify the syscall agent. The
    same agent implementation is used across different sandboxes created with
    \coolname{}. Our framework only provides interfaces to change how the loader
    and the monitor interact with the agent.}} The full sandbox development
workflow is shown in~\Cref{fig:workflow}.

\subsection{Use Cases}
\label{sec:usecases}

We apply the aforementioned development workflow to build PKU-based sandboxes
for ERIM and XOM-Switch. We describe the sandbox's components (the monitor, the
loader and the syscall agent) below and their interactions for each use case.

\subsubsection{Use Case 1 -- A Sandbox for ERIM}\par\hfill
\label{sec:usecase1}

The PKU-based sandbox intercepts and monitors the following system calls:
\texttt{modify\_ldt}, \texttt{prctl}, \texttt{seccomp}, \texttt{ptrace},
\texttt{process\_vm\_readv}, \texttt{process\_vm\_writev}, \texttt{mprotect},
\texttt{pkey\_mprotect}, \texttt{pkey\_alloc}, \texttt{pkey\_free},
\texttt{mmap}, \texttt{munmap}, \texttt{mremap}, \texttt{execve},
\texttt{shmat}, \texttt{shmdt}, all variants of the \texttt{fork/clone} system
calls, and all variants of the \texttt{open} system call. The loader adds these
system calls to the \texttt{SList} such that our system call agent reports them
to the monitor. We exclude \texttt{open}-like system calls from \texttt{SList},
however, because many applications invoke these system calls so frequently that
monitoring them from user space would add substantial run-time
overhead~\cite{connor2020pkupitfalls}. The loader also adds the \texttt{inodes}
of all \emph{sensitive} files\footnote{Currently, \texttt{IList} only contains
  the \texttt{inode} of \texttt{/proc/self/mem}. However, we can easily expand
  \texttt{IList} in case that we discover more sensitive files.} to
\texttt{IList}. Next, the loader passes these two lists to the syscall agent and
initializes it.

At run time, the agent forwards the system calls included in \texttt{SList} to
the monitor, while the rest are executed natively, bypassing \texttt{ptrace}. It
also intercepts \texttt{open}-like calls directly in kernel space and denies
opening of sensitive files to protect against attackers that aim to abuse
these files to bypass memory isolation
(see~\Cref{sec:inconsistencies}). Furthermore, the agent blocks opening of hard
and soft links to sensitive files, since they use the same \texttt{inodes}. The
monitor also ensures that the agent is only initialized \emph{once} after
\texttt{execve} or \texttt{fork/clone}-like system calls.

The monitor can inspect the \texttt{PKRU} register to reliably determine whether
the program is currently executing code in U or T. This is necessary to stop
certain attacks that rely on data races to bypass the sandbox
(see~\Cref{sec:races}). In addition, it tracks pages which are in
M\textsubscript{T} by intercepting \texttt{pkey\_mprotect} system calls from
T. An attacker that controls U can use system calls to access M\textsubscript{T}
directly or to unlock it to U
(see~\Cref{sec:inconsistencies,sec:influencing,sec:modtrustmapping}). To prevent
such attacks, our monitor forbids the following system calls from U:
\texttt{modify\_ldt}, \texttt{prctl} setting \texttt{seccomp}, \texttt{seccomp},
\texttt{ptrace}, \texttt{pkey\_alloc}, \texttt{pkey\_free},
\texttt{pkey\_mprotect}, \texttt{shmat} and \texttt{shmdt}. It also rejects
\texttt{process\_vm\_readv} and \texttt{process\_vm\_writev} system calls that
attempt to access M\textsubscript{T} from U.

The monitor scans all executable pages U loads into the address space at run
time for unsafe instructions. To do so, it intercepts and monitors
\texttt{mmap}, \texttt{mremap} and \texttt{mprotect} system calls from U as
these calls could introduce unsafe instructions or unlock
M\textsubscript{T}. The memory scanning process ignores our call gates and
\texttt{xrstor} instructions that are followed by proper checks. These sequences
are also considered safe by ERIM. We implemented a modified version of Hodor's
instruction vetting scheme in user space using \texttt{ptrace} to deal with the
discovered unsafe instructions. 

Unlike Hodor, however, our scheme vets \emph{both} unsafe \texttt{wrpkru} and
\texttt{xrstor} instructions. In addition, Hodor's sandbox always terminates
execution whenever a hardware breakpoint is triggered, while our monitor allows
execution of unsafe \texttt{xrstor} instructions to continue if bit 9 of eax
register is not set. If the bit \emph{is} set, we terminate the program. The
monitor also implements techniques described
in~\Cref{sec:relocation,sec:relocation2} to deal with attacks that use code
relocation to introduce unsafe instructions. Eliminating unsafe instructions
using binary rewriting reduces \emph{debug register pressure} and, thus,
decreases the likelihood that we have to resort to single-step execution when a
page contains more unsafe instructions than the number of debug registers
(see~\Cref{sec:hodor}).

Similar to previous work~\cite{connor2020pkupitfalls}, our monitor enforces
W\textsuperscript{$\wedge$}X by intercepting and monitoring system calls that
map pages and change permissions, since Linux by default allows pages that are
both writable and executable. Moreover, the monitor does not allow executable
mappings that are \texttt{MAP\_SHARED} or \texttt{MAP\_SHARED\_VALIDATE} to
protect against attackers that try to modify an immutable mapping via another
mutable mapping of the same shared memory (see~\Cref{sec:mappings}). Dealing
with attacks that try to directly modify the underlying object of a file-backed
mapping is more complicated, though. First, the monitor intercepts system calls
that map memory (e.g., \texttt{mmap}) and replaces the file-backed mappings,
with \texttt{MAP\_ANONYMOUS} ones, ensuring there are no mappings that are
backed by a file. Then the monitor copies the file contents that the application
initially attempted to map to the mapped region. To prevent attacks through
memory pages with mutable backing files (see~\Cref{sec:mappings}), the monitor
needs to impose restrictions on mapped regions, e.g., we reject any
attempt to map pages that are simultaneously \texttt{MAP\_SHARED} and
executable. Although we did not observe any compatibility or usability issues
that arose from these restrictions, we discuss their potential implications in~\Cref{sec:discussion}.

\paragraph*{\textbf{Protecting Multi-Threaded Programs}} Protecting multi-threaded
programs is a challenge because hardware breakpoints are only set in the
thread-local register context and because malicious threads could attempt to
modify executable code while it is being scanned for unsafe instructions. This
opens new possibilities for attacks that would not be possible in single-threaded
programs, as we explain in~\Cref{sec:incompleteupdate,sec:races}. To
counter these threats, we spawn a unique monitor thread for every application
thread we protect. Monitor threads that supervise threads of the same process
share data structures and \emph{always} enter a critical section when they scan
the application memory for unsafe instructions. Monitor threads must enter the
same critical section when an application thread attempts to execute a system
call that could change the contents of any of the pages that are being scanned
(e.g., \texttt{mremap}). This design avoids the problem of race conditions
during memory scanning.

Secondly, whenever a process spawns a thread for the first time (i.e., when the
program transitions from single-threaded to multi-threaded execution), the
monitors stop relying solely on hardware breakpoints for instruction vetting,
since updates to the set of breakpoints would not propagate beyond the currently
executing thread as described in~\Cref{sec:incompleteupdate}. Specifically, the
monitors \emph{only} use hardware breakpoints to vet the unsafe instructions
whose addresses were stored in the debug registers at the moment the process
started to use multiple threads. The monitor marks code pages containing unsafe
instructions, not already protected by debug registers, as non-executable. This
includes code pages that were mapped as non-executable by the monitor before
switching to multi-threaded execution, as well as new pages mapped by the
threads. Attempts to execute instructions from code pages that were marked as
non-executable by the monitor trigger a fault. When the monitor is notified of
faults on one of these pages, which it can determine by inspecting the
instruction pointer, it does not terminate execution immediately. First, the
monitor checks if the instruction that was about to get fetched is an unsafe
instruction or not. If it is an unsafe \texttt{wrpkru} instruction, the monitor
terminates execution, while if it is an unsafe \texttt{xrstor}, the monitor
terminates execution \emph{only} if bit 9 of eax is set. Otherwise, the
instruction is considered safe and the monitor uses an emulation engine for x86
instructions (included in \coolname{}) to emulate the instruction, essentially
updating the program's state to make it seem like the instruction was
\emph{actually} executed.

At run time, the monitor can decide if emulation of instructions is necessary by
detecting if multiple threads are used. To do so, it intercepts system calls
that create and destroy threads, and checks the \texttt{/proc/self/task} file.
Even though emulating instructions in user space is slow, we did not experience
significant performance degradation in our experiments that included
multi-threaded applications, since unsafe instructions are rare as described in
previous work~\cite{hedayati2019hodor, vahldiek2018erim}. In addition,
eliminating a portion of unsafe instructions with SBI decreases the number of
instructions that should be emulated, since the monitor needs to mark fewer code
pages as non-executable. We provide details for the emulation engine
in~\Cref{sec:implementation} and we discuss alternative ways to deal with
multi-threaded applications in~\Cref{sec:discussion}.

Hodor implements memory isolation in a way similar to ERIM. As a result, we can
apply the sandbox we described above to Hodor as well, requiring only minor
modifications, e.g., swapping call gates.

\begin{table*}[t]
	\resizebox{1.0\linewidth}{!}{
		\begin{tabular}{|l | c | c | c | c|}
			\hline
			& \textbf{ERIM-CPI} & \textbf{ERIM-CPI with} & \textbf{ERIM-SS} & \textbf{ERIM-SS with}\\
			\textbf{APP} & \textbf{(No Sandbox)} & \textbf{\coolname{} Sandbox} & \textbf{(No Sandbox)} & \textbf{\coolname{} Sandbox}\\
			\hline
			\textbf{\texttt{nginx} (1 worker)}					& 6.54\% 	& 4.97\%	& 0.57\% & 1.91\%\\
			\hline
			\textbf{\texttt{nginx} (2 workers)}					& 5.72\% 	& 6.20\%	& 4.62\% & 3.32\%\\
			\hline
			\textbf{\texttt{nginx} (3 workers)}					& 1.55\% 	& 2.23\%	& -1.22\% & 1.05\%\\
			\hline
			\textbf{\texttt{lighttpd} (1 worker)}				& -- 	& --	& 0.16\% & 2.14\%\\
			\hline
			\textbf{\texttt{lighttpd} (2 workers)}				& -- 	& --	& 0.22\% & 0.23\%\\
			\hline
			\textbf{\texttt{lighttpd} (3 workers)}				& -- 	& --	& -0.01\% & -0.03\%\\
			\hline
			\hline
			\textbf{geometric mean}				& 3.87\% 	& 4.10\%	& 0.71\% & 1.43\%\\
			\hline
		\end{tabular}
	}
    \caption{We isolated shadow stacks and safe regions in CPI/CPS with ERIM
      (ERIM-SS and ERIM-CPI respectively). We measured the overhead of
      standalone ERIM-SS and ERIM-CPI when they are not protected by a sandbox
      (No Sandbox) compared to the native execution. We report the overhead of
      ERIM-SS and ERIM-CPI with \coolname{} sandbox compared to the native
      execution. \coolname{} sandbox here refers to the sandbox described
      in~\Cref{sec:usecase1}. -- indicates that the experiment failed.}    
	\label{tab:saferegionoverhead}
\end{table*}

\subsubsection{Use Case 2 -- A Sandbox for XOM-Switch}\par\hfill
\label{sec:usecase2}

We also used \coolname{} to build a sandbox for XOM-Switch. To the best of our
knowledge, this is the first PKU-based sandbox for XOM-Switch. This sandbox
prevents attackers from abusing PKU to disable XOM, and, thus, eliminates the
requirement of an additional defense like CFI to protect XOM
(see~\Cref{sec:xom}).

The XOM-Switch sandbox is similar to the sandbox described above. Therefore, we
focus only on their implementation differences. First, we treat XOM-Switch as a
PKU-based memory isolation scheme, in which U includes all the application
code. As a result, the sandbox does not need to reliably determine the current
domain (it is always U). Second, M\textsubscript{T} consists of the eXecute Only
Memory (XOM). Since XOM-Switch uses the enhanced version of \texttt{mprotect},
the monitor tracks which pages are in M\textsubscript{T} by inspecting the
arguments of \texttt{mprotect} calls from U that apply XOM. Third, the monitor
considers all \texttt{wrpkru} instructions unsafe, since inter-domain
transitions are performed by the kernel and XOM-Switch does not rely on user
space \texttt{wrpkru} instructions for inter-domain transitions
(see~\Cref{sec:xom} for details). Last, the monitor denies any permission
changes of XOM; if a page is marked as execute only, it remains like that for
the entire execution.

%-------------------------------------------------------------------------------
\section{Implementation Details}
%-------------------------------------------------------------------------------
\label{sec:implementation}

\coolname{} consists of the following components: an SBI tool, the \coolname{}
monitor, the \coolname{} loader, the \coolname{} APIs, and syscall agent
(see~\Cref{fig:workflow}). We used ERIM's binary rewriter as the SBI tool. We
implemented the \coolname{} monitor, \coolname{} loader and \coolname{} APIs in
9536 lines of code. The emulation engine is part of the provided APIs and
currently supports 170 x86 instructions. We also implemented the syscall agent
with a small kernel patch (79 lines of code) for Linux kernel 5.3.18. The
\coolname{} monitor and \coolname{} loader are designed to be extensible. We
extended them using the \coolname{} APIs to implement PKU-based sandboxes for
ERIM and XOM-Switch. These extensions consist of 1007 lines of code. The
sandboxes we constructed have only minor implementation differences, as
mentioned in~\Cref{sec:usecase2}. We plan to open source our framework,
evaluation scripts, and proof-of-concept attacks upon acceptance of this paper.

%-------------------------------------------------------------------------------
\section{Evaluation}
%-------------------------------------------------------------------------------

We evaluated the performance and the security of the PKU-based sandboxes created
with \coolname{}.

\subsection{Performance}

We ran our experiments on an HP Z6 G4 workstation with a 12-core Intel Xeon
Silver 4214 CPU running at 2.20 GHz and 64 GB of RAM (Turbo-Boost and
Hyper-Threading were disabled). The machine runs Ubuntu 18.04.6 LTS with version
5.3.18 of the Linux kernel. We applied a minimal kernel patch that implements
the syscall agent. We evaluated the constructed PKU-based sandboxes on popular
high performance server applications: \texttt{nginx}, \texttt{lighttpd} and
\texttt{redis}. We ran a benchmarking client on a separate machine that is
connected to the workstation through a gigabit Ethernet connection. The client
machine has a 6-core Intel Core i7-8700K CPU running at 3.70 GHz and 64 GB of
RAM (Turbo-Boost and Hyper-Threading were disabled). The client machine runs
Ubuntu 18.04.6 LTS with version 5.4.0 of the Linux kernel. For \texttt{nginx}
and \texttt{lighttpd}, we used \texttt{wrk} benchmark to request a 4KB page for
10 seconds over 10 concurrent connections. For \texttt{redis}, we used
\texttt{redis-benchmark}, distributed with \texttt{Redis}, with the default
workload (100000 requests and 50 parallel connections). 

For the experiments described in~\Cref{sec:isolatingkeys}, the client
communicates with the server over HTTPS, while for the experiments described
in~\Cref{sec:protxom,sec:saferegions} communication happens over HTTP. We
measured the throughput of the server applications running under our defense
relative to the throughput of the native execution. We ran each experiment 10
times, removed the highest and lowest values as outliers, and reported the
average of the 8 remaining values. We configured \texttt{lighttpd} and
\texttt{nginx} to use 1--3 workers, and \texttt{redis} to use 1--3 I/O
threads. The server applications can saturate the network connection when
configured with 3 workers (\texttt{lighttpd}, \texttt{nginx}) and 3 I/O threads
(\texttt{redis}). As a result, we did not try configurations with more than 3
workers and I/O threads.

The developed PKU-based sandboxes, described in~\Cref{sec:usecases}, identified
unsafe instructions in \texttt{nginx}, \texttt{redis}, \texttt{ld.so},
\texttt{libm.so} and \texttt{libc.so} during our experiments. We managed to
eliminate an unsafe \texttt{xrstor} instruction in \texttt{nginx} with ERIM's
SBI tool, but we could not neutralize the other unsafe instructions with this
tool. We did not investigate further the reason that the tool failed, since the
constructed sandboxes do not solely rely on the SBI being successful in removing
all unsafe instructions.

\begin{table}[t]
	\resizebox{0.8\linewidth}{!}{
		\begin{tabular}{|l | c|}
			\hline
			\multirow{2}{*}{\textbf{APP}} & \textbf{ERIM-OpenSSL with}\\
			& \textbf{\coolname{} Sandbox} \\
			\hline
			\textbf{\texttt{nginx} (1 worker)}					& 1.12\%	\\
			\hline
			\textbf{\texttt{nginx} (2 workers)}					& 0.66\%	\\
			\hline
			\textbf{\texttt{nginx} (3 workers)}					& 1.21\%	\\
			\hline
			\textbf{\texttt{lighttpd} (1 worker)}				& 0.44\%	\\
			\hline
			\textbf{\texttt{lighttpd} (2 workers)}				& 0.31\%	\\
			\hline
			\textbf{\texttt{lighttpd} (3 workers)}				& 0.49\%	\\
			\hline
			\textbf{\texttt{redis} (1 I/O thread)}	& -1.29\%	\\
			\hline
			\textbf{\texttt{redis} (2 I/O threads)}	& 1.34\%	\\
			\hline
			\textbf{\texttt{redis} (3 I/O threads)}	& 0.01\%	\\
			\hline
			\hline
			\textbf{geometric mean}	& 0.47\%	\\
			\hline
		\end{tabular}
	}
	\caption{We isolated OpenSSL keys in server applications with ERIM
		(ERIM-OpenSSL). We report the overhead of ERIM-OpenSSL with \coolname{}
		sandbox compared to the native execution. \coolname{} sandbox here refers to
		the sandbox described in~\Cref{sec:usecase1}.}
	\label{tab:keyisolation}
\end{table}

\subsubsection{Protecting safe regions in CPI/CPS and SS}\par\hfill
\label{sec:saferegions}

Similar to previous work~\cite{vahldiek2018erim}, we used ERIM to isolate safe
regions of CPI/CPS~\cite{kuznetsov.etal+14}. We changed 13 lines of code to fix
an LLVM bug and to port the CPI compiler of ERIM to our testing environment
(Ubuntu 18.04 with kernel 5.3.18). In the same manner, we used ERIM to isolate
the safe regions of a shadow stack implementation (SS for
short)~\cite{burow2019sok}. To do so, we added 38 lines of code to the SS
compiler passes to add ERIM's functionality. We refer to the above compiler
passes as ERIM-CPI and ERIM-SS respectively. We applied ERIM-CPI to
\texttt{nginx} and ERIM-SS to \texttt{lighttpd} and \texttt{nginx}. We could not
run \texttt{lighttpd} after applying ERIM-CPI to it because of CPI's imprecise
handling of aliasing relations between memory references. We also verified that
\texttt{lighttpd} fails with the original CPI
compiler~\cite{kuznetsov.etal+14}. Similarly, we could not run \texttt{redis}
after applying either ERIM-CPI or ERIM-SS. Again, we verified that
\texttt{redis} also fails after compilation with the original
CPI~\cite{kuznetsov.etal+14} and SS~\cite{burow2019sok} compiler. Consequently,
we concluded that it is not our code that is responsible for the failures.

We show the overhead for each successful experiment
in~\Cref{tab:saferegionoverhead}. For the experiments, we removed ERIM's sandbox
(no sandbox) or replaced it with the sandbox described in~\Cref{sec:usecase1}
(\coolname{} sandbox). For ERIM-CPI with \coolname{} sandbox we report overhead
of 2.23--6.20\% with geometric mean of 4.10\%, while for ERIM-SS with
\coolname{} sandbox we report overhead of -0.03--3.32\% with geometric mean of
1.43\%. We also measured the overhead of standalone ERIM-CPI and ERIM-SS without
the protection of any sandbox to show that even in the worst case (6.20\%), most
of the overhead (5.72\%) can be attributed to the PKU-based memory isolation
scheme and not the sandbox.

\subsubsection{Isolating OpenSSL keys in server applications}\par\hfill
\label{sec:isolatingkeys}

Similar to previous work~\cite{vahldiek2018erim}, we isolated OpenSSL session
keys in popular server applications with ERIM (ERIM-OpenSSL), to protect against
server application vulnerabilities such as
Heartbleed~\cite{10.1145/2663716.2663755}. We configured \texttt{lighttpd},
\texttt{nginx} and \texttt{redis}, through their config files, to use
ERIM-OpenSSL and only use ECDHE-RSA-AES128-GCM-SHA256 cipher and AES encryption
for sessions. For the experiments, we replaced ERIM's sandbox with the sandbox
described in~\Cref{sec:usecase1} (\coolname{} sandbox). Our results are shown
in~\Cref{tab:keyisolation}. We report an overhead of -1.29--1.34\% with
geometric mean of 0.47\%.

\begin{table}[t]
	\resizebox{0.8\linewidth}{!}{
		\begin{tabular}{|l | c|}
				\hline
				\multirow{2}{*}{\textbf{APP}} & \textbf{XOM-Switch with} \\
				     & \textbf{\coolname{} Sandbox} \\
				\hline
				\textbf{\texttt{nginx} (1 worker)}					& 0.04\%	\\
				\hline
				\textbf{\texttt{nginx} (2 workers)}					& -0.02\%	\\
				\hline
				\textbf{\texttt{nginx} (3 workers)}					& -0.09\%	\\
				\hline
				\textbf{\texttt{lighttpd} (1 worker)}				& 0.02\%	\\
				\hline
				\textbf{\texttt{lighttpd} (2 workers)}				& 0.50\%	\\
				\hline
				\textbf{\texttt{lighttpd} (3 workers)}				& 0.16\%	\\
				\hline
				\textbf{\texttt{redis} (1 I/O thread)}	& 1.48\%	\\
				\hline
				\textbf{\texttt{redis} (2 I/O threads)}	& 0.88\%	\\
				\hline
				\textbf{\texttt{redis} (3 I/O threads)}	& 0.00\%	\\
				\hline
				\hline
				\textbf{geometric mean}	& 0.33\%	\\
				\hline
			\end{tabular}
		}
  \caption{We applied eXecute Only Memory (XOM) using XOM-Switch. We protected
    XOM-Switch with the \coolname{} sandbox. \coolname{} sandbox here refers to
    the sandbox described in~\Cref{sec:usecase2}. We report the overhead of
    XOM-Switch with this sandbox compared to the native execution.}
	\label{tab:xomswitchoverhead}
\end{table}

\begin{table*}[t]
	\centering
	\resizebox{0.7\linewidth}{!}{
		\begin{tabular}{|l | c | c | c|}
			\hline
			\multirow{2}{*}{\textbf{Sandbox Characteristics}} & \textbf{Hodor's} & \textbf{ERIM's} & \textbf{\coolname{}}\\
			& \textbf{Sandbox} & \textbf{Sandbox} & \textbf{Sandboxes}\\
			\hline
			\textbf{Handling of unsafe instructions}		& Incomplete	& Incomplete & Complete  \\
			\hline
			\textbf{Kernel Modifications}					& Major 	& Minor	& Minor\\
			\hline
			\textbf{PKU Pitfalls Protection}				& \xmark 	& \xmark	& \cmark$^*$\\
			\hline
			\textbf{New PKU Pitfalls Protection}			& \xmark 	& N/A & \cmark \\
			\hline
			\textbf{Performance overhead}					& Low 	& Low	& Low\\
			\hline
		\end{tabular}
	}
	\smallskip
	\resizebox{0.3\linewidth}{!}{$^*$Except for signal context attacks (see~\Cref{sec:signals}).}
	\caption{Comparison of different PKU-based sandboxes. We refer to the
      sandboxes described in~\Cref{sec:usecases} as \coolname{} sandboxes.}
	\label{tab:comparisontable}
\end{table*}

\subsubsection{Protecting Execute Only Memory}\par\hfill
\label{sec:protxom}

We used XOM-Switch~\cite{xom-switch} to apply eXecute Only Memory
(XOM). XOM-Switch is vulnerable to attackers that attempt to abuse PKU to
disable XOM, unless it is combined with an additional mitigation such as CFI. We
lift this requirement by combining XOM-Switch with the PKU-based sandbox
described in~\Cref{sec:usecase2} (\coolname{} sandbox). Our results are depicted
in~\Cref{tab:xomswitchoverhead}. We report overhead of -0.09--1.48\% with
geometric mean of 0.33\%.

\begin{table}[!t]
	\centering
	\resizebox{1.0\linewidth}{!}{
		\begin{tabular}{l|c c}
			\hline
			\multirow{2}{*}{\textbf{Attack}}   & \textbf{\coolname{}}\\
			& \textbf{Sandboxes}\\
			\hline
			\hline
			Inconsistencies of PT permissions~\cite{connor2020pkupitfalls} & \cmark\\\hline
			Inconsistencies of PKU permissions~\cite{connor2020pkupitfalls} & \cmark\\\hline
			Mapping with mutable backings~\cite{connor2020pkupitfalls} & \cmark\\\hline
			Changing code by relocation~\cite{connor2020pkupitfalls} & \cmark\\\hline
			Influencing intra-process behavior with seccomp~\cite{connor2020pkupitfalls} & \cmark\\\hline
			Modifying trusted mappings~\cite{connor2020pkupitfalls} & \cmark\\\hline
			Race conditions in scanning~\cite{connor2020pkupitfalls} & \cmark\\\hline
			Determination of trusted mappings~\cite{connor2020pkupitfalls} & \cmark\\\hline
			Signal context attacks~\cite{connor2020pkupitfalls} & \xmark\\\hline
			Vetted unsafe instruction relocation~\Cref{sec:relocation2} & \cmark\\\hline
			Incomplete debug register update~\Cref{sec:incompleteupdate} & \cmark\\\hline
			\hline

	\end{tabular}}
    \caption{Security analysis of the PKU-based sandboxes described
      in~\Cref{sec:usecases}. We refer to these sandboxes as \coolname{}
      sandboxes. \cmark{} \color{black} indicates that the sandbox stops the
      attack, while~\xmark{} \color{black} indicates the opposite.}
	\label{tab:securitytable}
\end{table}

\subsection{Security and Completeness}

We analyzed the security of the constructed sandboxes on existing
proof-of-concept attacks~\cite{connor2020pkupitfalls} and the two additional
attacks that we discovered while building \coolname{}
(see~\Cref{sec:newpitfalls}). For the former, we used the open source
implementation of the exploits\footnote{https://github.com/VolSec/pku-pitfalls},
while for the latter we used our proof-of-concept exploits. We verified that the
developed sandboxes stop all the attacks described
in~\Cref{sec:pitfalls,sec:newpitfalls}, except for signal context attacks
(see~\Cref{sec:signals}). This is not a fundamental limitation of our approach,
but of the current prototypes of the framework and the constructed sandboxes. We
discuss potential solutions to stop signal context attacks
in~\Cref{sec:discussion}. Our results are depicted in~\Cref{tab:securitytable}.

We compared the sandboxes described in~\Cref{sec:usecase1,sec:usecase2} (we
refer to them as \coolname{} sandboxes) with the ones provided by ERIM and
Hodor. The comparison is shown in~\Cref{tab:comparisontable}. \coolname{}
sandboxes are the first PKU-based sandboxes that can handle unsafe instructions
without causing security or usability issues
(see~\Cref{sec:inshandling}). \coolname{} sandboxes are also the first PKU-based
sandboxes that prevent the attacks described
in~\Cref{sec:pitfalls,sec:newpitfalls}, except for signal context attacks
(see~\Cref{sec:signals}). The new attacks described in~\Cref{sec:newpitfalls}
specifically target Hodor's instruction vetting mechanism, and are not
applicable (N/A) to ERIM. Both \coolname{} sandboxes and ERIM's sandbox are
implemented in user space; nevertheless they require a small kernel patch to
optimize performance. Last, all the sandboxes in this list incur low performance
overhead.

%-------------------------------------------------------------------------------
\section{Discussion}
%-------------------------------------------------------------------------------
\label{sec:discussion}

In this section, we discuss the limitations of our approach and alternative solutions.

\paragraph*{\textbf{Signal context attacks}} A concurrent work with
ours~\cite{bumjinendokernel}, proposes Endokernel, a subprocess virtualization
scheme to deal with challenges in in-process isolation. Specifically for
signals, the authors describe a signal virtualization mechanism that prevents
attackers from tampering \texttt{PKRU} by abusing signals
(see~\Cref{sec:signals}). However, two of the three provided implementations of
Endokernel rely on additional mitigations, software diversity and CFI, which
come with their own limitations regarding efficacy and performance.

Some of the techniques used in Endokernel could be applied to our framework to
prevent signal context attacks. We believe that this would be feasible, since
\texttt{ptrace} can intercept, delay, deny and redirect signals in ways similar
to Endokernel~\cite{bumjinendokernel}.

\paragraph*{\textbf{Restrictions on memory mappings}} \coolname{} restricts
memory mappings to deal with attackers that target mappings with mutable
backings as described in~\Cref{sec:usecase1}. This might lead to usability
issues in applications such as older JIT engines, which used double-mapping as a
way to bypass SELinux's W\textsuperscript{$\wedge$}X
policy~\cite{selinux,selinux2}. Modern JIT engines no longer use double-mapping, however, since it can be detrimental to the application's security~\cite{wxjitfirefox}.

\paragraph*{\textbf{Multi-threaded applications}} Currently, we use instruction
emulation to deal with applications that use multiple threads. However, this
might add significant overhead in cases where several pages contain unsafe
instructions. One suitable alternative could be to use process-wide events
monitoring that will be available in future
kernels~\cite{synchronousevents}. Otherwise, we could implement a
\emph{stop-the-world} mechanism using signals to ensure that all the threads of
a process are stopped, and then update their debug registers synchronously. This
would require though to first provide a concrete solution for signal context
attacks (see~\Cref{sec:signals}).

\paragraph*{\textbf{Multiple levels of trust}} We only experimented with
application partitioning schemes that use two levels of trust (trusted and
untrusted). However, with our framework it would be possible to develop
sandboxes for systems that use more than two memory domains, since the
constructed sandboxes can reliably determine the current executing domain by
inspecting \texttt{PKRU} as described in~\Cref{sec:design}, and track the
sensitive data of each domain.

%-------------------------------------------------------------------------------
\section{Related Work}
%-------------------------------------------------------------------------------

In-process isolation has been explored in depth, resulting in dozens of
systems. In this section, we summarize works on in-process isolation that do not
rely on PKU.

\paragraph*{\textbf{OS abstractions}} Previous work introduced OS abstractions
to enable multiple memory views and fast transition between them within a
process address space~\cite{chen2016shreds, litton2016lwc,
  10.5555/1387589.1387611,hsu2016smv}. These approaches expose thread-like
entities, control which resources they can access, and permit efficient
transitions between them. However, all these techniques are not directly
applicable to legacy code without code modifications.

\paragraph*{\textbf{Virtualization-based techniques}} Hodor~\cite{hedayati2019hodor}
and SeCage~\cite{10.1145/2810103.2813690} leverage virtualization
extensions~\cite{intel-mpk} (VT-x), to provide different memory views for
trusted and untrusted code. SeCage is vulnerable to an in-process adversary
unless it is used in conjunction with CFI, while Hodor's VT-x based
implementation is less efficient than Hodor's PKU-based
counterpart~\cite{hedayati2019hodor}. Intel and AMD CPUs provide Supervisor-mode
Access Prevention (SMAP) hardware feature to disable kernel accesses to user
space memory. Seimi uses SMAP and VT-x to provide low cost and
secure in-process isolation~\cite{9152611}. xMP~\cite{9152671} extended Xen hypervisor's
\texttt{altp2m} subsystem~\cite{10.1145/2664243.2664252,
  10.1145/3274694.3274698} and the Linux memory management system to isolate
sensitive user space and kernel data in disjoint xMP memory domains.

\paragraph*{\textbf{Hardware extensions}} Researchers also proposed hardware
extensions to provide efficient fine-grained component
isolation. CHERI~\cite{7163016} and CODOMs~\cite{6853202} extended the RISC and
x86 ISAs respectively, with capabilities. Donky~\cite{255298}, on the other
hand, augmented the x86 and RISC-V ISAs to provide secure memory protection
domains similar to PKU. MicroStache~\cite{mogosanu2018microstache} and
IMIX~\cite{217644} extended the x86 ISA with instructions to access safe
regions. ARM memory domains~\cite{arm-md} are similar to PKU domains, but they 
are only available on 32-bit chips and domain permissions can only be modified
with privileged instructions. This paper focuses on solutions that can be built
on commodity x86 CPUs.

\paragraph*{\textbf{SFI}} Software fault isolation (SFI)
restricts parts of an application code from accessing memory outside of
designated
bounds~\cite{10.1145/1629575.1629581,10.1145/168619.168635,yee2009native,donovan2010pnacl,10.5555/1404014.1404039,erlingsson2006xfi,zhao2011armor,deng2015isboxing,mccamant2006evaluating,267235}. SFI
techniques employ complex static and dynamic analysis and instrumentation that
introduce non-negligible overhead. In addition, many of the proposed techniques
rely on an additional mitigation such as CFI, to prevent in-process attackers
from bypassing bounds checks.

\paragraph*{\textbf{Compartmentalization}} Partitioning an application into
compartments and defining which resources they can access is an open problem and
it is orthogonal to this paper. Previous work focuses on identifying suitable
isolation boundaries in applications using automatic and semiautomatic (e.g.,
annotations) techniques~\cite{10.1145/3445814.3446728,brumley2004privtrans,
  gudka2015soaap,vasilakis2018breakapp,10.1145/3144555.3144563}. However,
completely automating compartmentalization of existing software is still
challenging.

%-------------------------------------------------------------------------------
\section{Conclusion}
%-------------------------------------------------------------------------------

Recent research has explicitly highlighted the extreme care that should be taken
when implementing PKU-based sandboxing, mentioning a large number of edge cases
and a difference in perspective between the OS and the security community on PKU
as contributing factors. In this paper, we analyzed the various challenges of
PKU-based sandboxing. We also introduced two new proof-of-concept attacks that
target Hodor and that manage to bypass its sandbox.

We then presented \coolname{}, a new PKU-based sandboxing framework that
facilitates development of PKU-based sandboxes. We applied our framework to
build sandboxes for two state-of-the-art PKU-based memory isolation systems:
ERIM and XOM-Switch. We evaluated the security and performance of the
constructed sandboxes using proof-of-concept exploits and high-performance
server applications respectively. Our extensive evaluation shows that
\coolname{} overcomes limitations of existing work, enabling practical,
efficient and secure PKU-based sandboxing.

%\nocite{*}

%%%%%%%%%%%%%%%%%%%%%%%%%%%%%%%%%%%%%%%%%%%%%%%%%%%%%%%%%%%%%%%%%%%%%%%%%%%%%%%%
{%\normalsize %\bibliographystyle{abbrv}
\bibliographystyle{ACM-Reference-Format}
\bibliography{garmr}}

\end{document}